\documentclass[dvips]{article}
\newcommand{\doublespacing}{\let\CS=\@currsize\renewcommand{\baselinesstrech}
{2.0}\tiny\CS}

\begin{document}

\textwidth 16cm
\newcommand{\bd}{\begin{document}}
\newcommand{\ed}{\end{document}}
\newcommand{\bc}{\begin{center}}
\newcommand{\ec}{\end{center}}
\newcommand{\bfr}{\begin{flushright}}
\newcommand{\efr}{\end{flushright}}
\newcommand{\lt}{\left}
\newcommand{\rt}{\right}
\newcommand{\vs}{\vspace}
\newcommand{\hs}{\hspace}
\newcommand{\beq}{\begin{equation}}
\newcommand{\eeq}{\end{equation}}
\newcommand{\lb}{\linebreak}
\newcommand{\pb}{\pagebreak}
\newcommand{\mb}{\makebox}
\newcommand{\fb}{\framebox}
\newcommand{\mc}{\multicolumn}
\newcommand{\ben}{\begin{enumerate}}
\newcommand{\een}{\end{enumerate}}
\newcommand{\bit}{\begin{itemize}}
\newcommand{\eit}{\end{itemize}}
\newcommand{\ol}{\overline}
\newcommand{\un}{\underline}
\newcommand{\lefq}{\lefteqn}
\newcommand{\ba}{\begin{array}}
\newcommand{\ea}{\end{array}}
\newcommand{\beqa}{\begin{eqnarray}}
\newcommand{\eeqa}{\end{eqnarray}}
\newcommand{\beqas}{\begin{eqnarray*}}
\newcommand{\eeqas}{\end{eqnarray*}}
\newcommand{\bfg}{\begin{figure}}
\newcommand{\efg}{\end{figure}}
\newcommand{\bds}{\begin{displaymath}}
\newcommand{\eds}{\end{displaymath}}
\newcommand{\btb}{\begin{tabbing}}
\newcommand{\etb}{\end{tabbing}}
\newcommand{\para}{\parallel}
\newcommand{\pad}{\partial}
\newcommand{\nn}{\nonumber}
\newcommand{\la}{\leftarrow}
\newcommand{\ra}{\rightarrow}
\newcommand{\lgla}{\longleftarrow}
\newcommand{\lgra}{\longrightarrow}
\newcommand{\La}{\Leftarrow}\newcommand{\Ra}{\Rightarrow}
\newcommand{\Lra}{\Leftrightarrow}
\newcommand{\Lgla}{\Longleftarrow}
\newcommand{\Lgra}{\Longrightarrow}
\newcommand{\bm}{\boldmath}
\newcommand{\lan}{\langle}
\newcommand{\ran}{\rangle}
\renewcommand{\a}{\alpha}
\renewcommand{\b}{\beta}
\newcommand{\g}{\gamma}
\newcommand{\G}{\Gamma}
\renewcommand{\d}{\delta}
\newcommand{\eps}{\epsilon}
\newcommand{\Th}{\Theta}
\newcommand{\s}{\sigma}
\newcommand{\lam}{\lambda}
\newcommand{\D}{\Delta}
\newcommand{\vare}{\varepsilon}
\newcommand{\pr}{\prime}
\newcommand{\ro}{\rho}
\newcommand{\nab}{\nabla}
\newcommand{\m}{\mu}
\newcommand{\n}{\nu}
\newcommand{\Sg}{\Sigma}
\newcommand{\p}{\pi}
\newcommand{\R}{I\!\!R}
\newcommand{\om}{\omega}
\newcommand{\Om}{\Omega}
\newcommand{\ze}{\zeta}
\newcommand{\vart}{\vartheta}
\newcommand{\tri}{\triangle}
\newcommand{\f}{\frac}
\newcommand{\iny}{\infty}
\newcommand{\pro}{\propto}
\bc {\huge Coherent states of non-Hermitian quantum systems} \ec

\vs{1cm}

\bc
{\it B. Roy{\footnote {e-mail : barnana@isical.ac.in}}\\
Physics \& Applied Mathematics Unit \\
Indian Statistical Institute \\
Kolkata - 700 108} \ec

\vs{.25cm}

\bc {and} \ec

\vs{.25cm}

\bc
{\it P. Roy{\footnote {e-mail : pinaki@isical.ac.in}} \\
Physics \& Applied Mathematics Unit \\
Indian Statistical Institute \\
Kolkata - 700 108} \ec

\vs{1cm}

\vs{1cm}

\bc {\large {\un{Abstract}}} \ec

We use the Gazeau-Klauder formalsim to construct coherent states of non-Hermitian quantum systems. In particular we use this formalism to construct coherent state of a $\cal{PT}$ symmetric system. We also describe the construction of coherent states following Klauder's minimal prescription.\\

\pb

\section{Introduction} In recent years non-Hermitian quantum mechanics have been extensively studied from various stand points \cite{moi}. Among the different non-Hermitian systems, there is a class of problems which are $\cal{PT}$ invariant, $\cal{P}$ and $\cal{T}$ being the parity and time reversal operator respectively and it was shown that some of the non-Hermitian $\cal{PT}$ symmetric problems admit real eigenvalues \cite{ben}. Subsequently it was pointed out that the reality of the spectrum is essentially due to $\eta$-pseudo Hermiticity \cite{mostafa1}. Interestingly some of the $\eta$-pseudo-Hermitian systems are also $\cal{PT}$ symmetric. Because of this property and also because of their intrinsic interest $\cal{PT}$ symmetric and $\eta$-pseudo-Hermitian potentials have been examined widely \cite{ben1,mostafa2}. 

On the other hand coherent states play an important role in the context of Hermitian quantum mechanics \cite{pere}. Recently the concept of coherent states was also introduced to $\cal{PT}$ symmetric quantum mechanics \cite{bagchi1}. However, as in Hermitian models, coherent states corresponding to arbitrary non-Hermitian potentials are not easy to construct. This is mainly due to the fact that symmetry of the problem i.e, a knowledge of the ladder operators may not always be known. This difficulty can however be overcome by using the Gazeau-Klauder (GK) approach \cite{klauder4}. Recently Klauder \cite{klauder} has suggested a set of requirements which a coherent state should satisfy and he proposed a method of constructing coherent states for solvable potentials. This method is simple and has been applied succesfully to a number of exactly solvable Hermitian potentials with discrete \cite{klauder1} as well as continuous spectrum \cite{klauder4}. Here our objective is to show that with an appropriately chosen inner product the GK formalism can also be extended to non-Hermitian potentials. In particular we shall use this technique to construct coherent states of the $\cal{PT}$ symmetric Scarf I potential. We shall also construct coherent states of the 
same potential satisfying a minimal set of requirements.

\section{Some results on $\cal{PT}$ symmetric systems}
Let us consider a Hamiltonian $H$ such that
\beq
H\psi_n(x) = E_n\psi_n(x) = \omega e_n\psi_n,~~  e_0 = 0\label{hamil1}
\eeq

Then $H$ is said to be $\cal{PT}$ symmetric if
\beq
H = {\cal{PT}}H{\cal{PT}}\label{pt}
\eeq
where 
\beq
\lt. \ba {lcl} {\cal{P}} ~ &:&
~ x \rightarrow -x, \qquad p \rightarrow -p  \\
{\cal{T}} ~ &:& ~ x \rightarrow x, ~~\qquad p \rightarrow -p, \qquad
i \rightarrow -i  \ea \rt\}
\eeq
Furthermore, if in addition to (\ref{pt}) the wave functions are also $\cal{PT}$ invariant i.e, 
\beq
{\cal{PT}}\psi_n = \pm \psi_n
\eeq
then $\cal{PT}$ symmetry is said to be unbroken. On the other hand if the Hamiltonian is $\cal{PT}$ invariant while the wave functions are not, then $\cal{PT}$ symmetry is said to be spontaneously broken. It may be noted that systems with spontaneously broken $\cal{PT}$ symmetry are characterised by complex conjugate pair of eigenvalues. Here we shall only consider systems with unbroken $\cal{PT}$ symmetry.
 
A major difference between the non-Hermitian theories and the Hermitian ones lies in the definition of inner product. In the case of $\cal{PT}$ symmetric systems neither the standard definition of inner product nor the straightforward generalisation, namely, 
\beq
\left<\psi_m|\psi_n\right>^{PT}=\int [{\cal{PT}}\psi_m]\psi_n dx = (-1)^n\delta_{mn}
\eeq
work because the norm becomes negative for some of the states. Consequently a modification is necessary so that the norm is always positive. Indeed it has been shown that for $\eta$-pseudo-Hermitian quantum mechanics it is possible to define an inner product which is positive definite i.e, the Hamiltonian is Hermitian with respect to that inner product \cite{mostafa1}. For $\cal{PT}$ symmetric cases a convenient option is to use the $\cal{CPT}$ norm.
For $\cal{PT}$ symmetric theories with unbroken $\cal{PT}$ symmetry the $\cal{CPT}$ inner product is defined by \cite{brody}
\beq
\left<\psi_m|\psi_n\right>^{{\cal{CPT}}}=\int [{\cal{CPT}}\psi_m]\psi_n dx=\delta_{mn}\label{norm}
\eeq
where $\cal{C}$ is called the charge operator and is defined by \cite{brody}
\beq
{\cal{C}}(x,y)=\sum_{n=0}^\infty \psi_n(x)\psi_n(y)
\eeq
The action of the charge operator on the eigenfunctions is given by
\beq
{\cal{C}}\psi_n(x)=\int {\cal{C}}(x,y)\psi_n(y)dy=(-1)^n\psi_n(x)
\eeq
Another property which would be very useful later is the completeness of eigenfunctions. In coordinate space the completeness property can be expressed in terms of the charge operator as
\beq
\sum_{n=0}^\infty \psi_n(x)[{\cal{CPT}}\psi_n(y)]=\sum_{n=0}^\infty (-1)^n\psi_n(x)\psi_n(y)=\delta(x-y)\label{unity}
\eeq

\section{GK formalism for $\cal{PT}$ symmetric systems}
It is known that coherent states can be constructed using different techniques (e.g, a coherent state may be defined as a minimum uncertainty state, annihilation operator eigenstate etc) and usually they have different properties. In the GK formalism a coherent state should satisfy the following criteria \cite{klauder4}:
\begin{enumerate}
\item Continuity of labelling 
\item Temporal stability 
\item Resolution of Identity
\item Action identity 
\end{enumerate}
 
For $\cal{PT}$ symmetric systems a GK coherent state is a two parameter state defined by \cite{klauder4,klauder}
\begin{equation}
\psi(x;J,\gamma) = \f{1}{\sqrt{{\cal{N}}(J)}} \sum_{n=0}^{\infty} \frac{J^\frac{n}{2}exp(-i\gamma e_n)}{\sqrt{\rho_n}}\psi_n(x) \label{coh}
\end{equation}
where $J\geq 0$ and $-\infty < \gamma < +\infty$ are two parameters and $\psi_n(x)$ are the eigenstates. In (\ref{coh}), $\rho_n$ are a set of numbers defined as 
\begin{equation}
\rho_0=1~~,~~\rho_n = \prod_{i=1}^{n}e_i \label{rho}
\end{equation}

It is now necessary to determine the normalisation constant ${\cal N}(J)$. Now using the $\cal{CPT}$ inner product (\ref{norm}) the normalisation constant ${\cal N}(J)$ can be obtained from the condition $\left<\psi(x;J,\gamma)|\psi(x;J,\gamma)\right>^{{\cal{CPT}}}=1$:
\begin{equation}
{\cal{N}}(J) = \sum_{n=0}^\infty \f{J^{n}}{\rho_n}~~,~~0 < J < R=\lim_{n\rightarrow \infty}(\rho_n)^{\frac{1}{n}} \label{norma}
\end{equation}

Let us now examine whether or not the coherent states (\ref{coh}) satisfy the criteria mentioned above. From the construction it is clear that the coherent states are continuous in their labels i.e, $(J,\gamma)\rightarrow (J',\gamma')
=> \psi(x;J,\gamma,x)\rightarrow \psi(x;J',\gamma')$. It may also be noted that although $H$ is not Hermitian, the time evolution operator is still given by $e^{-iHt}$. Using (\ref{hamil1}) and (\ref{coh}) the pseudo unitary time evolution is found to be
\beq
e^{-iHt}\psi(x;J,\gamma)=\psi(x;J,\gamma+\omega t)\label{temporal}
\eeq
From (\ref{temporal}) we find that the GK coherent states are temporally stable.
We now show that the coherent state also admits a resolution of unity:
\beq
\ba{lcl}
\int d\mu(\gamma,J)[{\cal{CPT}}\psi(y;J,\gamma)]f(J)\psi(x;J,\gamma)&=& {\displaystyle \lim_{\Gamma\rightarrow\infty}\frac{1}{2\Gamma}~\int_{-\Gamma}^{\Gamma}d\gamma\int_0^\infty [{\cal{CPT}}\psi(y;J,\gamma)]\psi(x;J,\gamma)~f(J)~dJ}\\\\
&=& \displaystyle \sum_{n=0}^\infty \frac{(-1)^n\psi_n(y)\psi_n(x)}{\rho_n}\int J^n  \f{f(J)}{{\cal{N}}(J)}dJ=\delta(x-y)
\ea
\eeq
if the following moment problem has a solution:
\begin{equation}
\rho_n = \int_0^\infty J^n \f{f(J)}{{\cal{N}}(J)}dJ\label{moment}
\end{equation}
Thus we find that the GK coherent state for $\cal{PT}$ symmetric systems provide a resolution of unity subject to the solution of the moment problem (\ref{moment}).

We now proceed to check the action identity. Using (\ref{norma}) it can be easily cheked that
\beq
\begin{array}{lcl} \left<\psi|H|\psi\right>^{CPT}&=&\displaystyle{\f{1}{{\cal{N}}(J)}\sum_{m,n=0}^\infty \displaystyle{\f{J^{(n+m)/2}e^{-i\gamma(e_n-e_m)} \omega e_n}{\sqrt{\rho_m\rho_n}}(-1)^n}\left<\psi_m|\psi_n\right>}\\\\
&=&\displaystyle{\f{1}{{\cal{N}}(J)}\sum_{m,n=0}^\infty \displaystyle{\f{J^{(n+m)/2}e^{-i\gamma(e_n-e_m)} \omega e_n}{\sqrt{\rho_m\rho_n}}}(-1)^{n+m}\delta_{mn}=\omega J}\\
\end{array}
\eeq

so that the criteria $(4)$ is also satisfied. Thus the coherent state (\ref{coh}) satisfy all the criteria $(1)-(4)$. It may be noted the construction ultimately boils down to a solution of the moment problem (\ref{moment}).

\section{GK coherent states for $\cal{PT}$ symmetric Scarf I potential} 
As an example we shall now construct GK coherent state of an exactly solvable $\cal{PT}$ symmetric potential. Thus we consider the $\cal{PT}$ symmetric Scarf I potential \cite{levai,znojil,levai1}. The Schr\"odinger equation is given by
\beq
\left[-\f{d^2}{dx^2}+V(x)\right]\psi_n=E_n\psi_n\label{scarf}
\eeq
where the potential $V(x)$ is given by
\beq 
V(x) = \frac{2(\alpha^2+\beta^2)-1}{4}\f{1}{cos^2(x)}+\frac{(\alpha^2-\beta^2)}{2}\frac{sin(x)}{cos^2(x)}-\frac{(\alpha+\beta+1)^2}{4}~~,~~x\in[-\frac{\pi}{2},\frac{\pi}{2}]\label{pto} 
\eeq 
where $\alpha$ and $\beta$ are complex parameters such that $\beta^*=\alpha$ and $\alpha_R>\f{1}{2}$.
The eigenvalues and the corresponding eigenfunctions are given by \cite{levai,znojil,levai1}
\beq 
E_{n} = \omega e_n= n(n+2\alpha_R+1) , \qquad n =
0,1,2, \cdots 
\eeq 
\beq \psi _{n} (x) = \displaystyle{N_n~(1-sinx)^{\frac{\alpha}{2}+\frac{1}{4}}(1+sinx)^{\frac{\beta}{2}+\frac{1}{4}}P_n^{(\alpha,\beta)}(sinx)} \label{wave}
\eeq  where $N_n$ denotes the normalisation constant and $P_n^{(a,b)}$ are Jacobi polynomials \cite{grads}. 
It may be noted that the Hamiltonian is $\cal{PT}$ symmetric as well as $\cal{P}$-pseudo-Hermitian:
\beq
\ba{lcl}
H &=& {\cal{PT}}H{\cal{PT}}\\
H^{\dagger}&=&{\cal{P}}^{-1}H\cal{P}\\
\ea
\eeq
Also the wave functions are $\cal{PT}$ invariant:
\beq
{\cal{PT}}\psi_n(x) = \psi_n(x)
\eeq
and they satisfy the relation 
\beq
\int [{\cal{CPT}}\psi_m(x)]\psi_n(x) dx = \delta_{mn}
\eeq
In this case 
\beq
e_n=n(n+\nu)~,~\rho_0=1~,~\rho_n = \frac{\Gamma(n+1)\Gamma(n+\nu+1)}{\Gamma(\nu+1)}~,~R=\infty \label{en}
\eeq
where $\nu=2\alpha_R+1$.

Thus the GK coherent state for the Scaf I potential (in the coordinate representation) is given by
\beq
\psi(x;J,\gamma) = \f{1}{\sqrt{{\cal{N}}(J)}} \sum_{n=0}^{\infty} \frac{J^\frac{n}{2}exp(-i\gamma e_n)}{\sqrt{\rho_n}}\psi_n(x) \label{cohosci}
\eeq
where $\psi_n(x)$ and $\rho_n$ are given by (\ref{wave}) and (\ref{en}) respectively. From (\ref{norma}) the normalisation constant is found to be
\beq
{\cal{N}}(J)=\left[\Gamma(\nu+1)\sum_{n=0}^\infty \f{J^n}{\Gamma(n+1)\Gamma(n+\nu+1)}\right] = J^{-\nu/2}\Gamma(\nu+1)I_\nu(2\sqrt J)
\eeq 
where $I_\nu(x)$ stands for the Bessel function of the first kind \cite{grads}.

It is now necessary to show that the moment problem (\ref{moment}) has a solution. Using the relation \cite{grads} 
\beq
\int_0^\infty x^\mu K_\delta(ax)dx = 2^{\mu-1}a^{-\mu-1}\Gamma\left(\frac{1+\mu+\delta}{2}\right)\Gamma\left(\frac{1+\mu-\delta}{2}\right)~,~~Re(\mu+1\pm \delta)>0,~Re(a)>0
\eeq
we find that 
\beq
f(J) = \frac{1}{\pi} I_\nu(2\sqrt J)K_\nu(2\sqrt J)>0
\eeq
provides a solution to the moment problem (\ref{moment}). Now it can be shown that
\beq
\ba{lcl}
\int d\mu(J,\gamma)~[{\cal{CPT}}\psi(y;J,\gamma)]f(J)\psi(x;J,\gamma) &=& \displaystyle{\f{1}{2\pi}}\int_0^{2\pi}d\gamma \int_0^\infty dJ ~{\cal{N}}(J)^{-1}\psi(x;J,\gamma)[{\cal{CPT}}\psi(y;J,\gamma)]\\\\
&=& {\displaystyle \sum_{n=0}^\infty (-1)^n\psi_n(x)\psi_n(y)}=\delta(x-y) \label{resol}
\ea
\eeq
Thus the coherent state provide a resolution of unity. It can now be easily shown that the states (\ref{cohosci}) have many other properties characteristic of coherent states e.g they are non orthogonal:
\beq
\left<\psi(x;J',\gamma')|\psi(x;J,\gamma)\right>^{CPT} = \f{\Gamma(\nu+1)}{\sqrt{{\cal{N}}(J){\cal{N}}(J')}}\sum_{n=0}^\infty \f{(JJ')^{n/2}}{\Gamma(n+1)\Gamma(n+\nu+1)}e^{i n(n+\nu)(\gamma-\gamma')}
\eeq

From the above example it is thus clear that subject to the solution of the moment problem (\ref{moment}), it is possible to construct GK coherent states of any exactly solvable $\cal{PT}$ symmetric or $\eta$-pseudo-Hermitian potential with real spectrum. 
\subsection{Minimal coherent state for non Hermitian potentials}
In the last section we considered coherent states satisfying four conditions. However, if we relax conditions $(2)$ and $(4)$ and construct coherent states following Klauder's minimal prescription \cite{klauder2}, then a wider class of such states can be generated \cite{klauder3}. In this case a coherent state is required to satisfy two of the four criteria, namely, (1) Continuity in labelling and, (2) Resolution of identity, mentioned earlier. Thus  a coherent state corresponding to (\ref{pto}) is defined as
\beq
\psi(x;\beta) = \f{1}{\sqrt{{\cal{N}}(|\beta|)}}\sum_{n=0}^\infty \f{\beta^n}{\sqrt{\rho_n}}\psi_n(x)\label{coh1}
\eeq
where $\beta=re^{i\theta}$ is a complex number and $\psi_n(x)$ are given by (\ref{wave}). Here $\rho_n$ are a set of positive constants which would be specified later. The normalisation constant is determined from the condition $\left<\psi(x;\beta)|\psi(x;\beta)\right>^{{\cal{CPT}}}=1$ and is given by
\beq
{\cal{N}}(|\beta|) = \sum_{n=0}^{\infty}\f{(\beta\bar \beta)^n}{\rho_n}\label{norm2}
\eeq
Clearly $\beta\rightarrow \beta'\Rightarrow \psi(x;\beta)\rightarrow \psi(x;\beta')$ i.e, the coherent state (\ref{coh1}) is continuous in the label. Furthermore 
\beq
\int d^2\beta [{\cal{CPT}}\psi(y;\beta)]f(|\beta|)\psi(x;\beta) = 2\pi \sum_{n=0}^\infty \f{(-1)^n\psi_n(y)\psi_n(x)}{\rho_n}\int_0^\infty r^{2n+1}\f{f(r)}{{\cal{N}}(r)}dr \label{resol2}
\eeq
Thus (\ref{coh1}) admits resolution of unity if the positive constants $\rho_n$ are such that the Stieltjes moment problem (\ref{moment}) has a solution i.e,
\beq
2\pi\int_0^\infty r^{2n+1}\f{f(r)}{{\cal{N}}(r)}dr = \rho_n
\eeq
In a recent work \cite{klauder3} solutions to such moment problems for a number of different forms of $\rho_n$ have been found by using Mellin transform technique. Here we consider a simple example and take $\rho(n) = \Gamma(2n+1)$. In this case we find from (\ref{norm2})
\beq
{\cal{N}}(r) = cosh(r)
\eeq
Now using the result 
\beq
\int_0^\infty x^{2n}e^{-x} dx = \Gamma(2n+1) 
\eeq
$f(r)$ is found to be
\beq
f(r) = \f{1}{2\pi}\f{e^{-r}}{r}
\eeq
Since $f(r)>0$, the coherent state 
\beq
\psi(x;\beta) = \displaystyle{\f{1}{\sqrt{cosh(r)}}}\sum_{n=0}^\infty \f{\beta^n}{\sqrt {\Gamma(2n+1)}}\psi_n(x)\label{coh2}
\eeq
with $\psi_n(x)$ are given by (\ref{wave}), admits a resolution of identity. Also it can be shown that overlap between two coherent states is
\beq
\left<\psi(x;\beta')|\psi(x;\beta)\right>^{\cal{CPT}} = \f{cosh(\sqrt{\beta'\beta})}{\sqrt{cosh(r')cosh(r)}}
\eeq
Thus (\ref{coh2}) furnishes a new coherent state of the $\cal{PT}$ symmetric Scarf I potential. We note that in contrast to the GK coherent state (\ref{cohosci}), the minimal coherent state (\ref{coh2}) is not temporally stable. 

\section{Conclusion}
Here it has been shown that by suitably modifying the normalisation procedure, it is possible to extend the GK formalism to non-Hermitian systems. Furthermore new families of coherent states may also be constructed following Klauder's minimal prescription. One such state with $\rho_n=\Gamma(2n+1)$ has been constructed here. Finally we note that the construction of coherent states outlined here may also be extended to $\eta$-pseudo Hermitian systems which are not necessarily $\cal{PT}$ symmetric \cite{ahmed10}.

\end{document}